# Proof of Directional Scaling Symmetry in Square and Triangular Lattices with the Concept of Gaussian and Eisenstein Integers

Cao Zexian

Institute of Physics, Chinese Academy of Sciences, Beijing 100190, China

**Abstract**    In a previous work [Scientific Reports 4, 6193(2014)] we proved the existence of scale symmetry in square and triangular (thus honeycomb) lattices by investigating the function $y = \arcsin(\sin(2\pi n x))$, where the parameter $x$ is either the silver ratio $\lambda = \sqrt{2} - 1$ or the platinum ratio $\mu = 2 - \sqrt{3}$. Here we give a new proof, simple and straightforward, by using the concept of Gaussian and Eisenstein integers. More importantly, it can be proven that there're infinitely many possibilities for scale symmetry in the square lattice, and one of them is even related to the golden ratio $\varphi = (\sqrt{5} - 1)/2$. The directions and the corresponding scale factors are explicitly specified. These results might inspire the search of scale symmetries in other even higher-dimensional structures, and be helpful for attacking physical problems modeled on the square and triangular lattices.

**I. Introduction**

Geometrical transformations include scaling, reflection, translation and rotation. Yet we noticed that in talking about the space groups of crystals, scaling is rarely mentioned[1]. Scale symmetry refers to size transformation that the resulting object has exactly the same properties as the original—It is notorious for the fact that it does not exist in most physical systems. When the dilation operation is limited to one unique direction, in this case we are dealing with the directional scaling symmetry, it is easily conceivable that even fewer objects can possess such a property. The expanding or compressing of a two-dimensional lattice, say, along an arbitrary direction usually brings with reordering of the neighborhood relationship among the lattice points, a messy situation for the lattice to resume the original structure.

The space groups for two-dimensional and three-dimensional lattices have since long been identified and applied to solving and understanding various physical problems in solid-state physics, surface physics, statistical physics, etc[1-2]. But figures or objects often have more than one line of symmetry. It is worth asking the question whether there is any directional scaling symmetry in the regular lattices, as any new symmetry can help understand the question from another perspective. In our previous publication[3], we proved the existence of directional scaling symmetry in both the square and the equilateral triangular (thus honeycomb) lattices by investigating the function $y = \arcsin(\sin(2\pi nx))$, where n is the integer variable and the parameter x is chosen to be either the silver ratio $\lambda = \sqrt{2} - 1$ or the platinum ratio $\mu = 2 - \sqrt{3}$, which can generate 8-fold and 12-fold quasiperiodic lattices at ease[4,5]. For each of the square lattice and the triangular lattice, a direction of scaling symmetry, together with the corresponding scale factor, has been explicitly determined. A natural question may be raised: Is such directional scaling symmetry unique or there are infinitely many possibilities? If there are infinitely many scale symmetries for these two-dimensional lattices, how can we find them out?

In the current work we will show that the concept of Gaussian integers and Eisenstein integers provide an easy path for the existence proof of scale symmetry in square lattice and triangular lattice, and for the former there are indeed infinitely many possibilities of scale symmetry. More surprisingly, both the silver ratio $\mu = 2 - \sqrt{3}$ and the golden ratio $\lambda = (\sqrt{5} - 1)/2$ are relevant to the scale symmetry of square lattice.

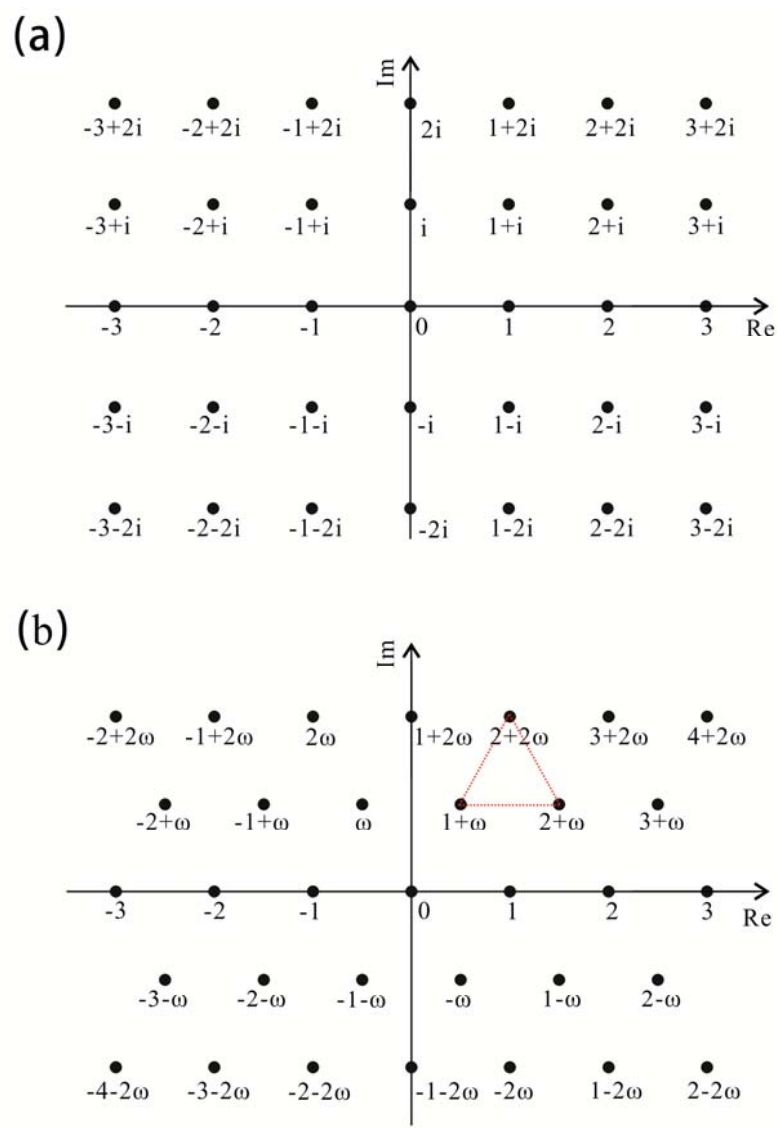

Figure 1. Square lattice (a) and equilateral triangular lattice (b) in a complex plane generated by the Gaussian integers and Eisenstein integers, respectively. The vortices of a unit triangle in (b) correspond to the Eisenstein integers $1+\omega$, $2+\omega$, and $2+2\omega$.

## II. Method

Geometrically, the Gaussian integers, i.e., complex numbers in the form of $m+in$ where both m and n are integer, form a square lattice (Fig.1a)[6]. Similarly, the Eisenstein integers, i.e., complex numbers in the form of $m+n\omega$ where both m and n are integer, and $\omega = (-1+i\sqrt{3})/2$ is one root of the equation $x^3 = 1$, form a triangular lattice (Fig.1b)[6]. With known parameters obtained by studying the graphs of the function $y = \arcsin(\sin(2\pi xn))$, the directional scaling symmetry for the square lattice and the triangular lattice can then be easily proven from another perspective. And more astonishing results concerning directional scaling symmetry will be revealed in the square lattice.

A Gaussian integer can be viewed as a vector in the complex plane $m+ni$, so is an Eisenstein integer $m+n\omega$, see Fig.1. Following Hamilton's binary representation of complex numbers, the Gaussian integers and Eisenstein integers can be equally treated as pairs of conventional integers $(m,n)$, which correspond to $m+ni$ in the former case, and to $m+n\omega$ in the latter case. We will use both notations interchangeably. The directional scaling symmetry is then referred to the transformation of the binary number $(m,n)$, which changes the scale along a peculiar direction and preserves the nature of the binary number, i.e., turn the integer pair $(m,n)$ in a one-to-one fashion, up to a scale factor, into another integer pair.

As pointed out in our previous publication[3], the drag point for the scaling transformation can be an arbitrary lattice point, thus without loss of generality, we consider the case that the drag point is chosen to be the origin (0, 0) in Fig.1. The direction of the scaling transformation is defined with regard to a side of the anchored unit cell, thus for the aim of proof it can be specified by the angle $\theta$ to the real axis. The directional scaling transformation can be formulated in two ways: (i) resolving the vector $(m,n)$ into parallel part and perpendicular part with regard to the direction of scaling transformation, scaling the parallel part by a factor $S_r$, and summing them up. In this case, the transformation can be represented as $(m, n) \mapsto S_r(m\cos\theta + n\sin\theta)(\cos\theta, \sin\theta) + (-m\sin\theta + n\cos\theta)(-\sin\theta, \cos\theta)$ ; (ii) rotating the vector $(m,n)$ by an angle of $-\theta$, scaling the real axis by a scale factor $S_r$, and then rotating the resulting vector by an angle of $\theta$. In the latter case, the transformation reads

$$m + ni \mapsto e^{i\theta} S_r e^{-i\theta}(m + ni) \qquad (1)$$

for square lattice, or

$$m + n\omega \mapsto e^{i\theta} S_r e^{-i\theta}(m + n\omega) \qquad (2)$$

for triangular lattice. The scale factor in the established directional scaling symmetry for square and triangular lattices is related to the angle $\theta$ via $S_r = \tan^2\theta$.[3]

## II. Results and discussion

### a. New proof of directional scaling symmetry

The directional scaling symmetry revealed in our previous work for the square lattice, which appears at a direction, taking an arbitrary lattice point as the drag point, at $\pi/8$ with respect to a side of the anchored unit square, the scale factor being $3 - 2\sqrt{2}$, can be rephrased, with the aid of Gaussian integer, as follows: The transformation

$$m + ni \mapsto e^{i\pi/8} S_r e^{-i\pi/8}(m + ni), \qquad (3)$$

where $S_r = 3 - 2\sqrt{2}$ denotes a scale change along the real axis, preserves the nature of Gaussian integers. By simple algebraic calculation one immediately finds that the transformation (1) turns the Gaussian integer $m + ni$ into another Gaussian integer $A + Bi = (m - n) + (3n - m)i$ up to a factor of $(2 - \sqrt{2})/2$. Since the transformation $(m,n) \mapsto (m - n, 3n - m)$ is linearly homogeneous, and of which the determinant of the coefficient matrix is positively definite, the Gaussian integers $(m - n) + (3n - m)i$ reproduce the square lattice. Q.E.D.

Similarly, the directional scaling symmetry for the triangular lattice, which appears at the direction, taking an arbitrary lattice point as the drag point, at $\pi/12$ with respect to a side of the anchored unit triangle, and the scale factor is $7-4\sqrt{3}$, can be rephrased, with the aid of Eisenstein integer, as follows: The transformation

$$m+n\omega \mapsto e^{i\pi/12} S_r e^{-i\pi/12}(m+n\omega) \quad, \tag{4}$$

where $S_r = 7-4\sqrt{3}$ denotes a scale change along the real axis, preserves the nature of Eisenstein integers. By simple algebraic calculation one immediately finds that the transformation turns the Eisenstein integer $m+n\omega$ into another Eisenstein integer $A+B\omega = n+(-m+4n)\omega$ up to a factor of $(2-\sqrt{3})$. Since the transformation $(m,n) \mapsto (n,-m+4n)$ is linearly homogenous, and the determinant of the coefficient matrix is positively definite, the Eisenstein integers $n+(-m+4n)\omega$ reproduce the triangular lattice. Q.E.D.

The one-to-one map for the transformation involved in the directional scaling symmetry is guaranteed by the fact that the physical operations including rotation and scale change don't alter the number of points or cause any point overlapping. So long as the resulting vector can be also represented as a Gaussian integer (Eisenstein integer) up to a scale factor, the square lattice (triangular lattice) is reproduced.

**b. Unveiling more directional scaling symmetries**

With the successful proof of directional scaling symmetry for the square lattice and equilateral triangular lattice illustrated above, one immediate question may be raised: Are there other directional scaling symmetries for these high-symmetry two-dimensional lattices? Let's first check out the square lattice.

We see that the directional scaling symmetry consists of two elements: an angle $\theta$ defining the direction, and the corresponding scale factor $S_r$. For any combinations of the directional angle $\theta$ and the scale factor $S_r$, they constitute a directional scaling symmetry for the square lattice so long as the general transformation (1) preserves the nature of the Gaussian integers. We will see that there are infinitely many such possibilities.

Starting from the transformation in (1), denoting $x = \tan\theta$, it has

$$e^{i\theta}S_r e^{-i\theta}(m+ni) = \{S_r(m+nx) - x(n-mx) + i[n-mx+xS_r(m+nx)]\}/(1+x^2) \quad (4)$$

As in the known case above, which is also desirable for the scale symmetry of a 2D lattice, we consider the case $S_r = \tan^2\theta = x^2$, then (4) turns into

$$e^{i\theta}S_r e^{-i\theta}(m+ni) = \{nx^3 - nx + 2mx^2 + i(mx^3 - mx + nx^4 + n)\}/(1+x^2) \quad (5)$$

Can we find a proper value of $x$ for the right-hand side of eq.(5) to be a Gaussian integer up to a constant factor?

Let's resort to the known case presented above to see if there are any useful hints and tips. Noticing the fact that $\tan\pi/8 = \sqrt{2}-1$ is the positive root of the equation $x^2 - 1 = -2x$ which is clearly a special case of the general equation

$$x^2 - 1 = -kx, \quad (6)$$

where k is a positive integer, this reminds us that the positive root of the general equation (6), $x = (\sqrt{k^2+4} - k)/2$, probably can enable transformation (5) to preserve the nature of Gaussian integers. With equation (6) at hand, some simple algebra turns (5) into

$$e^{i\theta}S_r e^{-i\theta}(m+ni) = \{-kn + 2m + i[-km + (k^2+2)n]\}x^2/(1+x^2) \quad . \quad (7)$$

This is to say that up to a factor of $x^2/(1+x^2)$, the Gaussian integers $(m,n)$ have been transformed into $\{-kn+2m, -km+(k^2+2)n\}$ which also represents a square lattice, since this transformation is linearly homogeneous, and the determinant of the coefficient matrix, $k^2+4$, is always positively definite. Hence, this proves that there are infinitely many scale symmetries for the square lattice: the direction is specified by an angle $\theta$, and the corresponding scale factor by $S_r = \tan^2\theta$, where $\tan\theta = (\sqrt{k^2+4} - k)/2$ for k=1, 2, 3…. For the case of $k=2$, $\tan\theta = \sqrt{2}-1$, $S_r = 3 - 2\sqrt{2}$, this reproduces the result in last paragraph.

Interestingly, for $k=1$, $\tan\theta = (\sqrt{5}-1)/2$ is the golden ratio $\varphi$. The corresponding transformation for the Gaussian integer is $(m,n) \mapsto (-n+2m, -m+3n)$. The golden ratio $\varphi$, besides appearing in places such as the Pythagorean triangle, the Fibonacci parastichous spirals of plants, the Penrose-tiling, the mass ratio of bound states in 1D Ising model[7], the critical fugacity of the hard-hexagon model[8], and even the maximum of Hardy's probability for quantum system of arbitrary finite dimension[9], is now found in the directional scaling symmetry of square lattice: the direction is at $\theta \sim 31.7°$ ($\tan\theta = \varphi$) with regard to a side of the anchored unit square, and the scale factor is $S_r = \varphi^2 = (3-\sqrt{5})/2$.

Although directional scaling symmetry has been discovered in quite the same way for the equilateral triangular lattice as for the square lattice in our previous work[3], it seems, however, that the scheme used above to reveal more scale symmetries for the square lattice does not apply to equilateral triangular lattice. At the moment no new directional scaling symmetry can be established for the equilateral triangular lattice by the effort to find a proper transformation that preserves the nature of the Eisenstein integer. This may be reasonable as the equilateral triangular lattice *per se* has a lower symmetry.

Considering the notorious nonexistence of scale symmetry in most systems, the fact that there are infinitely many directional scaling symmetries for the square lattice is quite surprising. Yet we don't know whether the scale symmetry for the square lattice has been exhausted or not in such a way. Scale symmetry is, to the best knowledge of the author, still not well understood. Its implications to physical problems defined on such high-symmetry two-dimensional lattice, for instance the Ising model and Hubbard model which are essentially important for condensed matter physics, deserve the attention of our theoretical colleagues.

**IV. Summary**

In summary, by using the concept of Gaussian integers and Eisenstein integers we proved the existence of directional scaling symmetry for the square lattice and the equilateral triangular lattice, and for the square lattice there are infinitely many possibilities. The directions and the corresponding scale factors for the scaling transformation are explicitly determined. What it may imply for physical problems that show a fundamental symmetry of square lattice or triangular lattice deserves more meticulous research. We wish the results here will catch the attention of theoretical physicists.